\documentstyle[preprint,aps]{revtex}
\newcommand{\sumv}{\sum_v}
\newcommand{\tr}[1]{\mbox{\rm tr} \left\{ #1 \right\}}
\newcommand{\Lm}{\gamma_\mu (1-\gamma_5)}
\begin{document}
\draft
\preprint{}
\title{Model independent determination of $|V_{ub}|$\\
       in heavy meson effective theory}
\author{Noriaki Kitazawa
 \thanks{Fellow of the Japan Society for the Promotion of Science
         for Japanese Junior Scientists.}
 \thanks{E-mail: kitazawa@theory.kek.jp}}
\address{National Laboratory for High Energy Physics (KEK),\\
         Tsukuba, Ibaraki 305, Japan
 \thanks{From October:
         Department of Physics, Tokyo Metropolitan University,
         Tokyo 192-03, Japan.}}
\date{\today}
\maketitle
\begin{abstract}
We propose a strategy
 of the model-independent determination
 of the Kobayashi-Maskawa matrix element $|V_{ub}|$
 within the framework of the heavy meson effective theory.
The effective theory is model-independent,
 since no assumption on the dynamics is introduced
 except for the symmetry of QCD.
Although the effective theory
 can only be applied to the decay process with soft pions,
 high detection efficiency on the soft pions in the future B-factories
 makes this theory effective.
The two exclusive processes,
 $B \rightarrow \pi l \nu$ and $B \rightarrow \pi \pi l \nu$,
 are used to determine the value of $|V_{ub}|$,
 under the condition that
 the values of the decay constants of $B$ and $B^*$ mesons are given.
\end{abstract}
%\pacs{}
\newpage

Precise measurement of the Kobayashi-Maskawa matrix elements
 is one of the approach to detect the new physics beyond the standard model
 at B-factories.
The unitarity relation of the matrix in the $B^0$-${\bar {B^0}}$ system,
\begin{equation}
 \sum_{i=u,c,t} V_{ib}^* V_{id} = 0,
\end{equation}
 is described as a triangle in the complex plane.
If the measured values of the sides and angles
 are not consistent with the triangle,
 the contribution from the new physics should exist.

Various strategies to measure the sides and angles are proposed.
Focusing on the sides,
 $|V_{cb}|$ in $|V_{cb}^* V_{cd}|$
 can be model-independently determined with enough accuracy
 by virtue of the heavy quark symmetry
 \cite{HQ-symmetry-1,HQ-symmetry-2,HQ-symmetry-3}.
But the determinations of the other two sides,
 $|V_{tb}^* V_{td}| \simeq |V_{td}|$ and
 $|V_{ub}^* V_{ud}| \simeq |V_{ub}|$,
 are still suffered from the ambiguity
 due to the theoretical estimation of the hadron matrix elements.

At present,
 $|V_{ub}|$ is extracted as a ratio $|V_{ub} / V_{cb}|$
 from the inclusive semi-leptonic B decay\cite{Vub/Vcb}.
The measured spectrum of the charged lepton momentum
 is understood as the combination of the two components of
 $B \rightarrow c l \nu$ and $B \rightarrow u l \nu$,
 which are proportional to $|V_{cb}|^2$ and $|V_{ub}|^2$, respectively.
The theoretical predictions
 of both spectra and decay rates of these two components
 are necessary to extract $|V_{ub} / V_{cb}|$.
There are some theoretical models
 \cite{ISGW,BSW,ACCMM}
 to calculate the hadron form factors with different predictions,
 and we have almost no criteria to select one of them at present.
The extracted values of $|V_{ub} / V_{cb}|$ are model dependent,
 and the ambiguity is larger than the present experimental error.
The other strategy which is free from the model dependence is expected.

Heavy meson effective theory\cite{heavy-meson} is not the model of the hadrons,
 since no assumption on the dynamics is introduced
 except for the symmetry of QCD.
Spin-flavor symmetry of the heavy quarks
 and the chiral symmetry of the light quarks restrict the interactions
 between the heavy-light mesons ($D$, $D^*$, $B$, and $B^*$)
 and the light pseudo-scalar mesons ($\pi$, $K$, and $\eta$),
 and the effect of the QCD dynamics
 is represented by the coupling constants of the interactions.
Once the coupling constants are fixed by using the experimental data,
 we can predict the form factors in model-independent way.
The form factors are effective only for the soft pions,
 since the chiral expansion is carried out
 in the construction of the effective theory.
In this letter,
 we propose a model-independent strategy of extracting $|V_{ub}|$
 through the fixing of the parameters of the exclusive processes
 $B \rightarrow \pi l \nu$ and $B \rightarrow \pi \pi l \nu$ with soft pions,
 under the condition that
 the values of the decay constants of $B$ and $B^*$ mesons are given.

The Lagrangian of heavy meson effective theory have been written down
 including the breaking effect of the spin-flavor symmetry
 \cite{HMET-1/M-1,HMET-1/M-2}.
Up to ${\cal O}(p^2)$ in the chiral expansion
 and ${\cal O}(1/M_Q^2)$ in the $1/M_Q$ expansion ($M_Q$ denotes the quark
mass),
\begin{eqnarray}
 {\cal L} &=&
 - \sumv \tr{ {\bar H}_v v \cdot iD H_v }
 - \sumv \tr{ {\bar H}_v {{(iD)^2} \over {2M}} H_v }
\nonumber\\
&&
 + \Lambda \sumv \tr{ {\bar H}_v H_v }
 + \kappa' \Lambda \sumv \tr{ {\bar H}_v {\Lambda \over M} H_v }
 + \kappa \Lambda \sumv
   \tr{ {\bar H}_v {\Lambda \over M} \sigma_{\rho\sigma}
                                 H_v \sigma^{\rho\sigma}}
\nonumber\\
&&
 + r \sumv \tr{ {\bar H}_v H_v v \cdot {\hat \alpha}_\parallel }
\nonumber\\
&& \qquad
 + r \sumv \tr{ {\bar H}_v {{iD_\mu} \over {2M}}
                H_v \alpha_\parallel^\mu } + h.c.
\nonumber\\
&& \qquad
 + r_1 \sumv
  \tr{ {\bar H}_v {\Lambda \over M} H_v
       v \cdot {\hat \alpha}_{\parallel} }
 + r_2 \sumv
  \tr{ {\bar H}_v {\Lambda \over M} \sigma^{\rho\sigma} H_v
       \sigma_{\rho\sigma} v \cdot {\hat \alpha}_{\parallel} }
\nonumber\\
&&
 + \lambda \sumv \tr{ {\bar H}_v H_v
                      \gamma_\mu \gamma_5 \alpha_\perp^\mu }
\nonumber\\
&& \qquad
 - \lambda \sumv \tr{ {\bar H}_v {{v \cdot iD} \over {2M}} H_v
                      \gamma_\mu \gamma_5 \alpha_\perp^\mu } + h.c.
\nonumber\\
&& \qquad
 - \lambda \sumv \varepsilon^{\mu\nu\rho\sigma}
  \tr{ {\bar H}_v {{iD_\rho} \over {4M}} H_v
       \sigma_{\mu\nu} \alpha_{\perp\sigma} }
 + \mbox{h.c.}
\nonumber\\
&& \qquad
 + \lambda_1 \sumv
  \tr{ {\bar H}_v {\Lambda \over M} H_v
       \gamma_\rho \gamma_5 \alpha_\perp^\rho }
 + \lambda_2 \sumv
  \tr{ {\bar H}_v {\Lambda \over M} \gamma_\rho \gamma_5 H_v
       \alpha_\perp^\rho }
\nonumber\\
&&
 + \mbox{(Anti-particle)},
\end{eqnarray}
 with $1/M = \mbox{diag}(1/M_c \ 1/M_b)$.
There are eight dimensionless parameters
 ($\kappa$, $\kappa$', $r$, $r_{1,2}$, $\lambda$, and $\lambda_{1,2}$)
 and a parameter $\Lambda$ with mass dimension.
The heavy mesons are described by the field
\begin{equation}
 H_v = {{1 + \not\!v} \over 2}
       \left[ i \gamma_5 P_v + \gamma_\mu P^{*\mu}_v \right],
\end{equation}
 where $v$ is the velocity of the heavy quark inside.
The fields $P_v$ and $P^*_v$
 are the heavy pseudo-scalar and heavy vector fields defined as
\begin{equation}
 P_v^{(*)} =
 \left(
  \begin{array}{ccc}
   D^0 & D^+        & D^+_s        \\
   B^- & {\bar B}^0 & {\bar B}^0_s
  \end{array}
 \right)^{(*)}
\end{equation}
 which have the mass dimension $3/2$.
The light pseudo-scalar mesons ($\pi$, $K$, and $\eta_8$)
 are described by the following fields as the Nambu-Goldstone bosons,
\begin{eqnarray}
 \alpha_\perp^\mu
  &=& {i \over 2} \left( \xi \partial^\mu \xi^{\dag}
                       - \xi^{\dag} \partial^\mu \xi \right),
\\
 \alpha_\parallel^\mu
  &=& {i \over 2} \left( \xi \partial^\mu \xi^{\dag}
                       + \xi^{\dag} \partial^\mu \xi \right),
\end{eqnarray}
 where
\begin{equation}
 \xi = e^{i\Pi/f_{\pi}},
\qquad \mbox{and} \qquad
 \Pi = \Pi^a {{\lambda^a} \over 2}.
\end{equation}
The light vector mesons ($\rho$, $K^*$, and $\omega_8$) are also introduced
 as the gauge bosons of the hidden local symmetry\cite{hidden-local}
 which is a model for the light vector mesons.
They are introduced
 through the covariant derivative of the heavy meson field
\begin{equation}
 i D_\mu H_v = i \partial_\mu H_v - H_v g_V V_\mu
\end{equation}
 and the combination
 ${\hat \alpha}_\parallel^\mu = \alpha_\parallel^\mu - g_V V^\mu$
 with $V_\mu = V^a_\mu \lambda^a/2$.
All the possible independent terms
 which are consistent with the symmetry
 (including $C$, $P$, and reparameterisation invariance)
 are enumerated in the Lagrangian.
The transformation properties of the fields
 and the details of the construction are described
 in the paper of ref.\cite{HMET-1/M-1}.

The heavy-to-light weak current
 can be written down in the same level of the expansion as
\begin{eqnarray}
 J_\mu^{ia}(0) &=& F \biggl[ \tr{ (\xi^{\dag})^{ji} \Lm H_v^{aj} }
\nonumber\\
&& \qquad\qquad
   + {1 \over {2M_a}}
     \tr{ (\xi^{\dag})^{ji} \Lm [\gamma_\rho, iD^\rho H_v^{aj}] }
  \biggr]
\nonumber\\
&&
 + \alpha_1 {\Lambda \over {M_a}}
   \tr{ (\xi^{\dag})^{ji} \Lm H_v^{aj} }
\nonumber\\
&&
 + \alpha_2 {\Lambda \over {M_a}}
   \tr{ (\xi^{\dag})^{ji} \Lm \gamma^\rho H_v^{aj} \gamma_\rho }
\nonumber\\
&&
 + \beta_1
   \tr{ (\xi^{\dag})^{ji} \Lm \left(H_v
        \left( v \cdot {\hat \alpha}_\parallel \right)\right)^{aj}}
\nonumber\\
&&
 + \beta_2
   \tr{ (\xi^{\dag})^{ji} \Lm \left(H_v
        \left( \gamma_\rho {\hat \alpha}_\parallel^\rho \right)\right)^{aj}}.
\end{eqnarray}
 with three parameters of the mass dimension $3/2$ ($F$ and $\alpha_{1,2}$)
 and two parameters of the mass dimension $1/2$ ($\beta_{1,2}$).
The decay constants of the heavy-light mesons are parametrized as
\begin{eqnarray}
 f_P &=& \sqrt{2 \over {M_Q}}
         \left\{ F + {\Lambda \over {M_Q}}
                ( \alpha_1 + 2 \alpha_2) \right\},
\\
 f_V &=& \sqrt{2 \over {M_Q}}
         \left\{ F + {\Lambda \over {M_Q}}
                ( \alpha_1 - 2 \alpha_2) \right\},
\end{eqnarray}
 where $f_P$ and $f_V$ are the decay constants
 of pseudo-scalar and vector heavy mesons, respectively
 \footnote{The QCD correction for the weak current
           is not considered for simplicity.}.

The Lagrangian gives the mass formulae for the heavy mesons
\begin{eqnarray}
 m_P^2 &=& M_Q^2 \left\{ 1 + 2 {\Lambda \over {M_Q}}
                           + 2 \kappa' {{\Lambda^2} \over {M_Q^2}}
                           + 12 \kappa {{\Lambda^2} \over {M_Q^2}}
                           + \omega_P {{\Lambda^3} \over {M_Q^3}}
                 \right\},
\\
 m_V^2 &=& M_Q^2 \left\{ 1 + 2 {\Lambda \over {M_Q}}
                           + 2 \kappa' {{\Lambda^2} \over {M_Q^2}}
                           - 4 \kappa {{\Lambda^2} \over {M_Q^2}}
                           + \omega_V {{\Lambda^3} \over {M_Q^3}}
                 \right\},
\end{eqnarray}
 where $m_P$ and $m_V$ are the masses
 of pseudo-scalar and vector heavy mesons, respectively.
The $1/M_Q^2$ correction to the un-squared masses $m_P$ and $m_V$
 are generally attached as the $\Lambda^3/M_Q^3$ terms
 with the coefficients $\omega_P$ and $\omega_V$ in the brackets.
Using a relation
\begin{equation}
 m_V^2 - m_P^2 = - 16 \kappa \Lambda^2
                 + (\omega_V - \omega_P) \Lambda^2 {\Lambda \over {M_Q}}
\end{equation}
 following from the above mass formulae,
 we can estimate the convergency of the $1/M_Q$ expansion.
The magnitude of the $1/M_Q^2$ correction is given as
\begin{equation}
 (\omega_V - \omega_P)
  \Lambda^2 \left( {\Lambda \over {M_c}} - {\Lambda \over {M_b}} \right)
 = (m_{D^{*0}}^2 - m_{D^0}^2) - (m_{B^{*0}}^2 - m_{B^0}^2)
 \simeq ( 396 \ \mbox{MeV} )^2,
\end{equation}
 where the values of the meson masses\cite{particle-data}
\begin{eqnarray}
 m_{D^0}    &\simeq& 1860 \ \mbox{MeV},
\\
 m_{D^{*0}} &\simeq& 2010 \ \mbox{MeV},
\\
 m_{B^0}    &\simeq& 5280 \ \mbox{MeV},
\\
 m_{B^{*0}} &\simeq& 5320 \ \mbox{MeV},
\end{eqnarray}
 are used.
If the quark masses are set as $M_c = 1500$MeV and $M_b = 5000$MeV
\begin{equation}
 (\omega_V - \omega_P) \Lambda^3 \simeq (695 \ \mbox{MeV})^3.
\end{equation}
Then we obtain the magnitude of the universal $1/M_Q$ correction
 to the squared mass difference as
\begin{equation}
 - 16 \kappa \Lambda^2 = (m_{D^{*0}}^2 - m_{D^0}^2)
                - (\omega_V - \omega_P) {{\Lambda^3} \over {M_c}}
                \simeq (597 \mbox{MeV})^2.
\end{equation}

We can see that the convergency of the $1/M_c$ expansion is not good,
 since the ${\cal O}(1/M_c)$ and ${\cal O}(1/M_c^2)$ corrections
 to the $m_D^2$ are the same order of magnitude.
The magnitude of the corrections relative to the leading are
\begin{eqnarray}
 \left| 12 \kappa {{\Lambda^2} \over {M_c^2}} \right| &\simeq& 0.12,
\\
 \left| \omega_P {{\Lambda^3} \over {M_c^3}} \right|
 &\simeq& \left| (\omega_V - \omega_P) {{\Lambda^3} \over {M_c^3}} \right|
  \simeq 0.10.
\end{eqnarray}
But the ${\cal O}(1/M_b^2)$ correction to $m_B^2$
 is one order of magnitude smaller than the ${\cal O}(1/M_b)$ correction:
\begin{eqnarray}
 \left| 12 \kappa {{\Lambda^2} \over {M_b^2}} \right|
  &\simeq& 1.1 \times 10^{-2},
\\
 \left| \omega_P {{\Lambda^3} \over {M_b^3}} \right|
 &\simeq& \left| (\omega_V - \omega_P) {{\Lambda^3} \over {M_b^3}} \right|
  \simeq 2.7 \times 10^{-3}.
\end{eqnarray}
This indicates that the convergency of the $1/M_b$ expansion is good.
Therefore,
 we apply this effective theory only to the B meson system in the following.

The decay amplitude of the process
 ${\bar B}^0 \rightarrow \pi^+ l {\bar \nu}$
 is obtained by calculating the diagrams of fig.\ref{pi}.
The form factors defined by
\begin{equation}
 \langle \pi(p_\pi) | J^{ub}_\mu (0) | {\bar B}^0 (p_B) \rangle
  = f_+(q^2) (p_B + p_\pi)_\mu + f_-(q^2) (p_B - p_\pi)_\mu
\end{equation}
 are obtained as
\begin{eqnarray}
 f_{\pm}(q^2)
 = {1 \over 2} {{f_B} \over {f_\pi}}
   \Bigg[
   1 &-& {{f_{B^*}} \over {f_B}}
        \left\{
        \lambda \left( 1 + {{v \cdot p_\pi} \over {2m_B}} \right)
        + \left( (\lambda_1 - \lambda_2) - \lambda \right) {\Lambda \over
{m_B}}
        \right\}
\nonumber \\
&&\qquad\qquad
 \cdot {{(m_B^2 - m_\pi^2 - q^2) \mp (m_B^2 - m_\pi^2 + q^2)}
                              \over {q^2 - m_{B^*}^2}}
   \Bigg],
\end{eqnarray}
 where $q^2 = (p_B - p_\pi)^2$ and
\begin{eqnarray}
 v \cdot p_\pi = {{m_B^2 + m_\pi^2 - q^2} \over {2m_B}}.
\end{eqnarray}
The explicit dependence on the bottom quark mass $M_b$
 is replaced by the dependence on the $B$ meson mass $m_B$
 up to the order of $1/M_b^2$ by using the mass formula.
The $B^*$ meson pole contribution emerges
 with the coupling constant $\lambda$
 and the combination $\lambda_1-\lambda_2$.
The mass of the pion
 is remained non-zero in the phase space integration only.
At the tree level,
 all the effect of the explicit chiral symmetry breaking
 due to the light quark masses, $m_u$ and $m_d$,
 is reduced to the isospin breaking effect which is negligibly small.
Then we can use the coupling constants of the chiral limit.

There are no contribution
 from the higher order terms of chiral expansion at the tree level.
Therefore,
 these form factors should be applied as far as the chiral expansion converges
 \footnote{This corresponds to the fact that
           only the ${\cal O}(p^2)$ and ${\cal O}(p^4)$ terms in chiral
expansion
           can contribute to the $\pi$-$\pi$ scattering amplitude
           at the tree level.}.
Following to the naive dimensional analysis\cite{naive},
 the chiral expansion converges if
\begin{equation}
 \left( {{E_\pi} \over {4 \pi f_\pi}} \right)^2 < 1,
\end{equation}
 where $E_\pi = v \cdot p_\pi$ denotes the energy of the pion
 in $B$ meson rest frame
 \footnote{Reparameterisation invariance always allow us to take $p_B = m_B
v$.}
 and $f_\pi \simeq 88$MeV denotes the decay constant of the pions
 in the chiral limit.
The $q^2$-spectrum
\begin{eqnarray}
 {{d\Gamma} \over {dq^2}}
 = {{G_F^2 |V_{ub}|^2} \over {24\pi^3}}
   \left\{ E_\pi^2 - m_\pi^2 \right\}^{3/2}
   |f_+(q^2)|^2,
\label{pi-spectrum}
\end{eqnarray}
 with the above $f_+(q^2)$ is effective in the region
\begin{equation}
 (m_B - m_\pi)^2 > q^2 > m_B^2 + m_\pi^2 - 2 m_B \cdot 4 \pi f_\pi,
\label{soft-region}
\end{equation}
 where the masses of the charged lepton and neutrino are neglected.

Once we fix the values of all the unknown parameters,
 $\lambda$, $((\lambda_1 - \lambda_2) - \lambda ) \Lambda / m_B$,
 $f_B$, and $f_{B^*}$,
 the value $|V_{ub}|$ can be obtained
 by fitting the spectrum of the above region.
If we consider the very soft pions, namely
\begin{equation}
 \left( {{E_\pi} \over {4 \pi f_\pi}} \right)^2 < 0.1
\label{soft-1}
\end{equation}
 or
\begin{equation}
 (m_B - m_\pi)^2 > q^2
                 > m_B^2 + m_\pi^2 - 2 m_B \cdot \sqrt{(4 \pi f_\pi)^2 / 10},
\end{equation}
 the $E_\pi / 2m_B$ dependence in the form factors
 can be approximately neglected in the $10\%$ level,
 and only the combination of the couplings,
 ${\bar \lambda}
   \equiv \lambda + ((\lambda_1 - \lambda_2) - \lambda ) \Lambda / m_B$,
 emerges in the form factor.

To fix the values of the coupling constants,
 we must consider the other independent processes of $B$ meson decay.
The decay amplitude of the process $B^- \rightarrow \pi^+ \pi^- l {\bar \mu}$
 is obtained by calculating the diagrams of fig.\ref{pi-pi}.
The form factors which are defined as\cite{Bl4-decay}
\begin{equation}
 \langle \pi^+(p_{\pi^+}) \pi^-(p_{\pi^-}) | J^{ub}_\mu (0)| B^-(p_B) \rangle
 = \omega_+ k_{+\mu} + \omega_- k_{-\mu} + r q_\mu
 + h i \epsilon_{\mu\alpha\beta\gamma} p_B^\alpha k_+^\beta k_-^\gamma,
\end{equation}
 where $k_\pm = p_{\pi^+} \pm p_{\pi^-}$ and $q = p_l + p_\nu$,
 are obtained as
\begin{eqnarray}
\omega_+ &=& - {i \over {2 f_\pi^2}} {{f_B} \over \sqrt{2}} \Bigg[ 1
   + {{f_{B^*}} \over {f_B}}
      \left\{
       \lambda
       \left( 1 + {{p_B \cdot k_+ - p_B \cdot k_-} \over {4m_B^2}} \right)
       + \left( (\lambda_1 - \lambda_2) - \lambda \right) {\Lambda \over {m_B}}
      \right\}
\nonumber\\
&&\qquad\qquad\qquad\qquad\qquad\qquad
\cdot
      {{2 m_B^2 - 3 \left( p_B \cdot k_+ - p_B \cdot k_- \right)}
       \over
       {m_B^2 - m_{B^*}^2 - p_B \cdot k_+ + p_B \cdot k_-}} \Bigg]
\nonumber\\
&& - {i \over {2 f_\pi^2}}
     \beta_1 \sqrt{m_B} \left( 1 - {\Lambda \over {2m_B}} \right)
     {{p_B \cdot k_-} \over {m_B^2}}
     {{k_+^2} \over {k_+^2 - m_\rho^2}},
\\
\omega_- &=& {i \over {2 f_\pi^2}} {{f_B} \over \sqrt{2}} \Bigg[ 1
   + {{f_{B^*}} \over {f_B}}
      \left\{
       \lambda
       \left( 1 + {{p_B \cdot k_+ - p_B \cdot k_-} \over {4m_B^2}} \right)
       + \left( (\lambda_1 - \lambda_2) - \lambda \right) {\Lambda \over {m_B}}
      \right\}
\nonumber\\
&&\qquad\qquad\qquad\qquad\qquad\qquad
\cdot
      {{2 m_B^2 - \left( p_B \cdot k_+ - p_B \cdot k_- \right)}
       \over
       {m_B^2 - m_{B^*}^2 - p_B \cdot k_+ + p_B \cdot k_-}} \Bigg]
\nonumber\\
&& - {i \over {2 f_\pi^2}}
     \left\{
      {{f_B} \over \sqrt{2}}
      - \beta_2 \sqrt{m_B} \left( 1 - {\Lambda \over {2m_B}} \right)
     \right\}
     {{k_+^2} \over {k_+^2 - m_\rho^2}},
\\
 h &=& {{i\sqrt{2}} \over {f_\pi^2}} f_{B^*}
       {1 \over {m_B^2 - m_{B^*}^2 - p_B \cdot k_+ + p_B \cdot k_-}}
       {1 \over {q^2 - m_{B^*}^2}}
      \left\{
       m_B^2 - {{p_B \cdot k_+ - p_B \cdot k_-} \over 2}
      \right\}
\nonumber\\
&&
\qquad
\cdot
      \lambda
      \Bigg[
       \lambda
       \left(
        1 + {{p_B \cdot k_+} \over {2 m_B^2}}
       \right)
       + 2 \left( \lambda_1 - \lambda \right) {\Lambda \over {m_B}}
       + 2 \lambda
            \left(
             {{p_B \cdot k_+ - p_B \cdot k_-} \over {2m_B^2}}
             - {\Lambda \over {m_B}}
            \right)
      \Bigg].
\end{eqnarray}
We do not write $r$, since it does not contribute
 if the masses of the lepton and neutrino are neglected
 \footnote{Three of the seven diagrams
           with the derivative coupling of the pseudo-scalar $B$ meson
           with $W$ bosons only contribute to $r$.}.
The relations
\begin{eqnarray}
 \left( \omega_+ - \omega_- \right)_{p_{\pi^+}=0}
 &=& - i {1 \over {\sqrt{2} f_\pi}} \cdot 2 f_+ ((p_B - p_{\pi^-})^2)
\\
 \left( \omega_+ + \omega_- \right)_{p_{\pi^-}=0}
 &=& 0
\end{eqnarray}
 which follow form the chiral symmetry,
 are satisfied.
The explicit dependence on the bottom quark mass $M_b$
 is replaced by the dependence on the $B$ meson mass $m_B$
 up to the order of $1/M_b^2$ by using the mass formula.
Some new unknown parameters emerge:
 $\beta_1$ and $\beta_2$ in the $\rho$ meson contribution
 in $\omega_+$ and $\omega_-$,
 and the new combination $\lambda_1 + \lambda_2$ in $h$.
Taking the S-wave configuration of pions in $B$ meson rest frame,
 the contributions of the $\rho$ meson and the form factor $h$ are dropped out.
The decay spectrum of the S-wave pions is given as
\begin{eqnarray}
 {{d\Gamma_{\rm S-wave}} \over {dk_+^2 dq^2}} &=&
 {{G_F^2 |V_{ub}|^2} \over {12 (4\pi)^5 m_B^3}}
 {{f_B^2} \over {f_\pi^4}}
 \left\{
  \left( {{m_B^2 - k_+^2 - q^2} \over 2} \right)^2 - q^2 k_+^2
 \right\}^{3/2}
\label{pi-pi-spectrum}
\\
&&
 \cdot
 \Bigg[
  1 + {{f_{B^*}} \over {f_B}}
      \left(
       \lambda
       + \left( (\lambda_1 - \lambda_2) - \lambda \right) {\Lambda \over {m_B}}
      \right)
  \left\{
   3 - {{4m_{B^*}^2} \over {2( m_{B^*}^2 - m_B^2 ) + m_B^2 + k_+^2 -q^2}}
  \right\}
 \Bigg]^2.
\nonumber
\end{eqnarray}
The coupling constants
 are included only in the combination of
 ${\bar \lambda}
  \equiv \lambda + ((\lambda_1 - \lambda_2)-\lambda) \Lambda / m_B$,
 since we are considering the very soft pions which satisfy
\begin{equation}
 \left( {{v \cdot k_+ / 2} \over {4 \pi f_\pi}} \right)^2 < 0.1,
\label{soft-2}
\end{equation}
 where $v \cdot k_+ / 2$ denotes the average of the pion energies
 in $B$ meson rest frame.
The $E_{\pi^-}/2m_B$ dependence in the form factors $\omega_+$ and $\omega_-$
 are approximately neglected in the $10\%$ level.
The next order terms of the chiral expansion (${\cal O}(p^2)$)
 can contribute to this process.
To keep the good approximation within $10\%$ error,
 we must set the condition of eq.(\ref{soft-2}).

The width of $D^{*+} \rightarrow D^0 \pi^+$ decay can be calculated as
\begin{equation}
 \Gamma(D^{*+} \rightarrow D^0 \pi^+)
 = {{m_D^2 \left( E_\pi^2 - m_\pi^2 \right)^{3/2}}
                          \over {12\pi m_{D^*}^2 f_\pi^2}}
   \left[ \lambda
        + \left((\lambda_1 - \lambda_2) - \lambda \right){\Lambda \over {m_D}}
   \right]^2,
\end{equation}
 where
\begin{eqnarray}
 E_\pi = {{m_{D^*}^2 - m_D^2 + m_\pi^2} \over {2m_{D^*}}}.
\nonumber
\end{eqnarray}
The experimental upper bound on this width\cite{D*->Dpi}
\begin{equation}
 \Gamma(D^{*+} \rightarrow D^0 \pi^+) < 131 \ \mbox{keV}
\end{equation}
 gives the upper bound on the parameter
\begin{equation}
 \lambda + \left((\lambda_1 - \lambda_2) - \lambda \right){\Lambda \over {m_D}}
  < 0.417.
\end{equation}
Although the $1/M_c$ expansion does not converge,
 this result gives the order of magnitude of the couplings.
We set ${\bar \lambda} = 0.4$ in the following order estimates.

We assume that the values of $f_B$ and $f_{B^*}$ are given.
The value of $f_B$
 will be obtained within $10\%$ accuracy
 by the lattice calculation in near future.
Although the value of $f_{B^*}$ is phenomenologically less important,
 \footnote{The weak decays of the $B^*$ meson are hard to observe,
           since the branching ratio is very small.}
 the lattice calculation of the value of $f_{B^*}$ can be interesting
 with a view to checking the convergency of the $1/M_b$ expansion.
We set $f_B=f_{B^*}=180$MeV in the following order estimates.

The decay rate of the ${\bar B}^0 \rightarrow \pi^+ l {\bar \nu}$
 with very soft pion is obtained
 by integrating the spectrum of eq.(\ref{pi-spectrum})
 over the phase space restricted by eq.(\ref{soft-1}).
The expected number of the events of such decay with $10^8$ $B$ mesons is
\begin{equation}
 \left. N({\bar B}^0 \rightarrow \pi^+ l {\bar \nu})
 \right|_{\rm vary \ soft \ pion} \simeq 2350 {{|V_{ub}|^2} \over {(1/20)^3}}.
\label{pi-number}
\end{equation}
The decay rate of the $B^- \rightarrow \pi^+ \pi^- l {\bar \nu}$
 with very soft pions is obtained in the same way
 by integrating the spectrum of eq.(\ref{pi-pi-spectrum})
 over the phase space restricted by eq.(\ref{soft-2}).
The expected number of the events with $10^8$ $B$ mesons is
\begin{equation}
 \left. N(B^- \rightarrow \pi^+ \pi^- l {\bar \nu})
 \right|_{\rm vary \ soft \ pion} \simeq 273 {{|V_{ub}|^2} \over {(1/20)^3}}.
\label{pi-pi-number}
\end{equation}
These number can be detectable
 in the future B-factories within the $10\%$ accuracy.
Therefore,
 it can be expected that the value of $|V_{ub}|$ is extracted
 by fitting both ${\bar \lambda}$ and $|V_{ub}|$ with these numbers.
The model-independent determination of $|V_{ub}|$ within $10\%$ accuracy
 can be expected.

We may obtain both the values of $\lambda$
 and $((\lambda_1 - \lambda_2) - \lambda) \Lambda/m_B$
 by fitting the spectrum of ${\bar B}^0 \rightarrow \pi^+ l {\bar \nu}$
 in the region of eq.(\ref{soft-region}).
The spectrum should be normalized
 by the number of eq.(\ref{pi-pi-number}) to factor out $|V_{ub}|$.
The convergency of the $1/M_b$ expansion can be estimated
 by comparing the values of $\lambda$
 and $((\lambda_1 - \lambda_2) - \lambda) \Lambda/m_B$.
Once these parameters are fixed,
 the value of $|V_{ub}|$ can be obtained by fitting the un-normalized
 spectrum of ${\bar B}^0 \rightarrow \pi^+ l {\bar \nu}$.
The extracted value is also the model-independent.

The accuracy of the extracted value
 depend on the experimental ability to detect the very soft pions,
 since the main theoretical ambiguity comes from the higher order contribution
 of the chiral expansion.
If we want the theoretical ambiguity of less than $5\%$,
 the phase space should be restricted as $(E_\pi / 4 \pi f_\pi)^2 < 0.05$
 in the process ${\bar B}^0 \rightarrow \pi^+ l {\bar \nu}$, for example.
The resultant number of the events will be very few and hard to detect.
If the precise measurement of the pure leptonic decays of $B$ mesons,
 $B^- \rightarrow \mu {\bar \nu}$ or $B^- \rightarrow \tau {\bar \nu}$,
 is easier, this strategy have no advantage.
Another large theoretical ambiguity
 may come form the $\pi$-$\pi$ re-scattering effect
 in $B^- \rightarrow \pi^+ \pi^- l {\bar \nu}$ decay.
But we can carry out the improvement
 by using the experimental data of the $\pi$-$\pi$ scattering phase shift.

In this letter
 we proposed a model-independent strategy
 to extract the value of $|V_{ub}|$ from the experiments.
Two kind of decay processes,
 ${\bar B}^0 \rightarrow \pi^+ l {\bar \nu}$ and
 $B^- \rightarrow \pi^+ \pi^- l {\bar \nu}$, with very soft pions are used.
The effect of the QCD dynamics
 is represented by only one parameter except for the decay constants,
 since the phase space of the decays is restricted
 so as to include only the very soft pions.
The values of the decay constants, $f_B$ and $f_{B^*}$,
 are expected to be obtained within the enough accuracy
 by the future lattice calculations.
The theoretical ambiguity can be less than $10\%$
 depending on the experimental ability to detect the soft pions,
 if the decay constants of $B$ and $B^*$ are known in good accuracy.

I am grateful to
 Yasuhiro Okada, Minoru Tanaka, and Masaharu Tanabashi
 of KEK theory group and Takeshi Kurimoto of Osaka University
 for fruitful discussions and comments.
I also would like to thank S.Uno of KEK
 for the helpful advice from the experimental point of view.

\begin{figure}
\caption{
The diagrams for ${\bar B}^0 \rightarrow \pi^+ l {\bar \nu}$ decay.
Black circle denotes the strong vertex,
 and black square denotes the weak vertex.
}
\label{pi}
\end{figure}
\begin{figure}
\caption{
The diagrams for $B^- \rightarrow \pi^+ \pi^- l {\bar \nu}$ decay.
Black circle denotes the strong vertex,
 and black square denotes the weak vertex.
}
\label{pi-pi}
\end{figure}

\end{document}